\documentclass[12pt]{article}
\usepackage{amsmath}
\usepackage{bm}
\usepackage{color}
\usepackage{caption}
\usepackage{subcaption}
\usepackage{graphicx}
\usepackage{braket}

\begin{document}
\title{Observation of spin-1 tunneling on a quantum computer}

\author{
Kh. P. Gnatenko \footnote{khrystyna.gnatenko@gmail.com},
V. M. Tkachuk\footnote{voltkachuk@gmail.com}\\
 Professor Ivan Vakarchuk Department for Theoretical Physics,\\
Ivan Franko National University of Lviv,\\
12, Drahomanov St., Lviv, 79005, Ukraine.}

\maketitle

\begin{abstract}
Spin-1 tunneling and splitting of energy levels as a result of tunneling are observed explicitly on IBM's quantum computer, ibmq-bogota. The spin-1 is realized with two spins-1/2.  We detect oscillations of spin-1 between the states $|1\rangle$, $|-1\rangle$ in the result of tunneling on the basis of studies of the time dependence of the mean value of z-component of spin-1 on the quantum device. The energy level splitting is observed quantifying on the IBM's quantum computer the eigenvalues of Hamiltonian which describes the spin tunneling.


\end{abstract}

\section{Introduction}

Quantum computers are natural devises for simulating quantum systems
\cite{Fau21,Adam19,Nic20}. Quantum bit in native way represents spin-1/2. Therefore  quantum
spin-1/2 systems or in general systems
with binary degrees of freedom are
the most suitable for simulating them with quantum processor \cite{Mon21,Her14,Gustafson,Kubra,Cervera-Lierta,Gna22,Gna21Probe,Gna21,Susulovska,Kuz21,Kuz20Let}.

Quantum spin tunneling is phenomena where single spin tunnels between two opposite
directions. This leads to degeneracy of energy levels related with opposite states
which is called quantum spin tunneling splitting (see, for instance, \cite{Kri17,Galetti}).
Experimental observation of quantum tunneling of the magnetization of cluster with $S=10$
was reported in \cite{Tho96}.
Latter in \cite{Tsu09,Kha10}
a direct measurement of the quantum tunneling splitting energy of the spin $S=1$ was done. Note that quantum spin tunneling splitting for zero field is only possible for integer spins. The smallest spin for which this phenomena can be observed is $S=1$.

In this paper we simulate phenomena spin-1 quantum tunneling on quantum computer.
We observe explicitly spin-1 tunneling and splitting of energy level as result of tunneling. The spin tunnelling is detected on the basis of studies of evolution of mean value of z-component of spin-1. The splitting of energy level is observed detecting energy levels of the Hamiltonian that describes single-spin tunneling on a quantum computer. The studies are done using the method of detection of the energy levels of spin system on a quantum computer with probe spin evolution proposed in \cite{Gna21Probe,Gna22}.

The paper is organized as follows. In Section 2 the spin-1 tunnelling is considered and the way to detect it on a quantum device is presented. In Section 3 the spin tunneling and the energy splitting in the result of tunneling are detected on IBM's quantum computer. Conclusions are presented in Section 4.

\section{Spin-1 tunneling and its studies on a quantum computer}
The Hamiltonian that describes single-spin tunneling reads
\begin{eqnarray}\label{HS1}
H=DS_z^2+\gamma (S_x^2-S_y^2),
\end{eqnarray}
here $S_{\alpha}$ are spin-1 operators, $D$ is the axial constant which
determines the magnetic anisotropy, constant $\gamma$ is responsible for single-spin tunneling effect (see, for instance, \cite{Fernandez}).

Energy levels and corresponding eigenstates of  (\ref{HS1})  are well known. They are as follows
\begin{eqnarray}
E=0, \ \ |\psi\rangle=\ket{0},\\
E=E_{\pm}=D\pm|\gamma|, \
|\psi_{\pm}\rangle={1\over\sqrt 2}(|1\rangle\pm|-1\rangle).
\end{eqnarray}
Tunneling leads to the splitting of energy level, it reads $\Delta=2|\gamma|$.

The  processes  of tunneling can be seen explicitly studying dynamical properties of spin-1.
In relation with this it is worth mentioning paper \cite{Fry08} where dynamical problems of spin-1 were considered to study quantum brachistochrone problem.

Let in the initial time $t=0$ the spin is in the state
$|1\rangle$ and is positively directed along z-axis, $\langle S_z\rangle_{t=0}=1$.  One can  find that the evolution of the state vector reads
\begin{eqnarray}
|\psi(t)\rangle=e^{-\frac{iHt}{\hbar}}|1\rangle=\cos\omega t |1\rangle-i\sin\omega t |-1\rangle,\label{mean}
\end{eqnarray}
where  $\omega=\gamma /\hbar$.
From (\ref{mean}) we have, that in result of tunneling the spin oscillates between two opposite directions described by the state vectors $|1\rangle$, $|-1\rangle$. These oscillations are reflected in the time dependence of the mean value of z-component of spin
\begin{eqnarray}
\langle S_z(t)\rangle=\langle\psi(t) |S_z|\psi(t)\rangle=\cos 2\omega t,
\end{eqnarray}
and can be detected on a quantum device.

Spin-1 can be realized with two spins-1/2 (see \cite{Sin03}).
The operator of spin-1 can be represented as a sum of two spin-1/2 operators  as follows
\begin{eqnarray}\label{S2s}
S_{\alpha}={1\over 2}(\sigma^{\alpha}_1+\sigma^{\alpha}_2),
\end{eqnarray}
where  $\sigma^{\alpha}_i$ are Pauli operators, $\alpha=(x,y,z)$. For spin-1 the eigenvalue of $S^2$ is $j(j+1)=2$, ($j=1$). In order to satisfy this relation  the
action of spin-1/2 operators has to be restricted on subspace spanned by vectors
\begin{eqnarray} \label{States2}
|00\rangle\equiv|1\rangle, \  |11\rangle\equiv|-1\rangle, \
{1\over\sqrt 2}(|01\rangle+|10\rangle)\equiv|0\rangle,
\end{eqnarray}
(see \cite{Sin03}).
Note that singled state of two spins-1/2 is annihilated by operators (\ref{S2s}) and does not belong to this subspace.
Representation of spin-1 by two spins-1/2 allows us to model and study spin-1 systems on a quantum computer, in particular to examine quantum spin-1 tunneling with quantum calculations.

In the next section we present results of simulation of spin-1 tunnelling on the IBM's quantum computer.

\section{Detection of spin-$1$ tunneling on IBM's quantum computer}

Using representation for spin-1 (\ref{S2s}) we rewrite Hamiltonian (\ref{HS1}) as follows
\begin{eqnarray}
H={D\over 2}(1+\sigma_1^z\sigma_2^z)+{\gamma\over 2}(\sigma_1^x\sigma_2^x-\sigma_1^y\sigma_2^y).\label{hhh}
\end{eqnarray}
Expression (\ref{hhh}) corresponds to Hamiltonian of two spins-$1/2$ with anisotropic Heisenberg interaction. Evolution operator of this system can be realized on a quantum computer.
Due to commutation relation $[\sigma_1^i\sigma_2^i,\sigma_1^j\sigma_2^j]=0$, the    evolution operator can be factorized. It reads
\begin{eqnarray}
U(t)=e^{-iHt}=e^{-i {D\over 2}t}e^{-i{D\over2}t\sigma_1^z\sigma_2^z}
e^{-i{\gamma\over2} t\sigma_1^x\sigma_2^x}e^{i{\gamma\over2}t\sigma_1^y\sigma_2^y}.
\end{eqnarray}
 Here for convenience we  put $\hbar=1$.

\begin{figure}[!!h]
		\centering
	\includegraphics[scale=0.6]{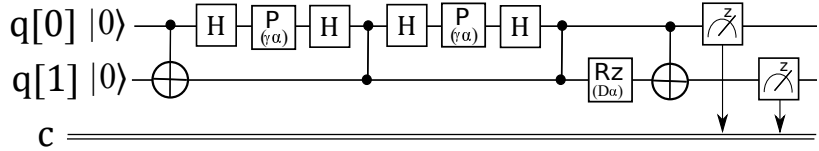}
		\caption{ Quantum protocol for studies of evolution of $\langle S_z(t)\rangle$ governed by Hamiltonian (\ref{HS1}), $\alpha=t$.}
		\label{fig:4}
	\end{figure}

In order to show quantum tunneling of spin-$1$ explicitly, we detect
evolution of the mean value
\begin{eqnarray}
\langle S_z(t)\rangle={1\over 2}(\langle\sigma^{z}_1(t)\rangle
+\langle\sigma^{z}_2(t)\rangle),
\end{eqnarray}
governed by (\ref{HS1}) on a quantum computer.
Quantum protocol for this studies is presented on
 Fig. \ref{fig:4}.

In the quantum protocol we consider the initial state of spin-1 to be $\ket{1}$, that corresponds to the state of two spins-1/2 (qubits) as $\ket{00}$. Also, to construct protocol Fig. \ref{fig:4} we  take into account that with the exactness to the total phase the operator $\exp(-i{\gamma\alpha}\sigma^x_0\sigma^x_1/2)$  can be represented as $CNOT_{01}H_0P_0(\gamma\alpha)H_0CNOT_{01}$, where $H_i$ is the Hadamard gate acting on $q[i]$, $CNOT_{ij}$ is the controlled NOT-gate acting on qubit $q[i]$ as on the controlled and on the qubit $q[j]$ as on the target. Operator $\exp(i{\gamma\alpha}\sigma^y_0\sigma^y_1/2)$ can be rewritten as $CNOT_{01}CZ_{01}H_0P_0(\gamma\alpha)H_0CZ_{01}CNOT_{01}$, where $P_0(\gamma\alpha)$ is the phase gate acting on qubit $q[0]$,  $CZ_{01}$ is the controlled Z-gate acting on qubits $q[0]$, $q[1]$. For $\exp(-i{D\alpha}\sigma^z_0\sigma^z_1/2)$ we have representation
$CNOT_{01}RZ_1(D\alpha)CNOT_{01}$, here  $RZ_1(D\alpha)$ is $Z$-rotation gate that acts on $q[1]$.  In quantum protocol Fig. \ref{fig:4} we also take into account that $(CNOT_{ij})^2=1$.

\begin{figure}[!!h]
		\centering
	\includegraphics[scale=0.3]{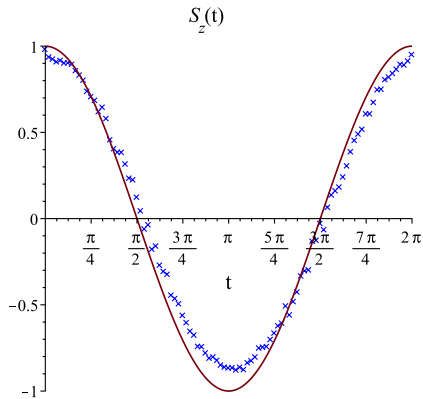}
		\caption{ Results of quantum calculations for the evolution of the mean value  $\langle S_z(t)\rangle$ governed by Hamiltonian (\ref{HS1}) with $D=-2$ and $\gamma=0.5$ on ibmq-bogota (marked by cresses) and analytical results (marked by line).}
		\label{fig:5}
	\end{figure}

We realized protocol Fig. \ref{fig:4} for $\alpha$ changing from $0$ to $2\pi$ with the step $\pi/48$ on ibmq-bogota. The results of quantum calculations  in the case of $D=-2$ and $\gamma=0.5$ are presented on  Fig. \ref{fig:5}.

On Fig. \ref{fig:5} we see that  during the time interval $[0, \pi]$ the spin-1 oscillates from the state $\ket{1}$ to $\ket{-1}$. During the next time interval  $[\pi, 2\pi]$ it returns to the initial state  $\ket{1}$. This reflects spin-1 tunnelling between two directions described by the state vectors $|1\rangle$, $|-1\rangle$.

\section{Detection of the energy spectrum splitting on a quantum computer}

To find the energy spectrum of spin-1 Hamiltonian  (\ref{HS1}) on a quantum device and detect its splitting we use the method of quantifying the energy levels of spin systems proposed in our papers \cite{Gna22,Gna21Probe}.
In \cite{Gna22} we presented a method for detecting the energy levels of a spin system which is based on the studies of evolution of the mean value of operator of a physical quantity anticommuting with the Hamiltonian of the system. Because an operator anticommuting with Hamiltonian does not exist for all spin systems,  in \cite{Gna21Probe} we generalized the proposed method of detecting energy levels to the case of arbitrary spin Hamiltonians.  In \cite{Gna21Probe} it was proposed to build the total Hamiltonian adding probe (ancila) spin-1/2 and to detect the energy levels on the basis of studies of the probe spin evolution.

In this section we apply the method of detection of the energy levels of spin systems  presented in \cite{Gna21Probe} to Hamiltonian  (\ref{HS1}) describing spin-1 tunneling problem.
So, let us construct the total Hamiltonian as follows
\begin{eqnarray}
H_T=\sigma^z_0(H+C),\label{total}
\end{eqnarray}
here $H$ is given by (\ref{HS1}),  $\sigma^z_0$ is $z$-component of the Pauli matrix corresponding to the probe spin-1/2. Constant $C$  has to be chosen to shift the energy levels of the Hamiltonian (\ref{HS1}) to the positive or negative ones. The energy levels of $H$   (\ref{HS1}) for $D<0$ and $\vert D\vert>\vert \gamma\vert$ are nonpositive. Therefore we can put $C=0$. In this case the energy levels of $H_T$ are related with the energy levels of the Hamiltonian (\ref{HS1}) as $E_T=\pm E$, where we use notation $E_T$ for the energy levels of $H_T$ and notation $E$ for the energy levels of $H$.

Operator $\sigma^x_0$ of the probe spin anticommutes with the total Hamiltonian. Let us study its evolution and detect the energy levels of $H_T$ and as result the energy levels of $H$ as was proposed in \cite{Gna22}.
We consider the initial state as
\begin{eqnarray}
|\psi_0\rangle=|+\rangle|\chi,\chi\rangle,\label{initial}
\end{eqnarray}
where  $|+\rangle={1\over\sqrt 2}(|0\rangle+|1\rangle)$ is the initial state of the probe spin. State $|\chi,\chi\rangle$ is the initial state of two spins-1/2 representing spin-1 with Hamiltonian (\ref{HS1}),
$|\chi\rangle={1\over\sqrt 2}(|0\rangle+e^{i\varphi}
|1\rangle)$. The state $|\chi,\chi\rangle$ belongs on the subspace (\ref{States2}). Note that state $|\chi,\chi\rangle$ with $\varphi\neq0$ includes all eigenstates of $H$.
Calculating  the evolution of the mean value of $\sigma^x_0$, we find
\begin{eqnarray}
\langle\sigma^x_0(t)\rangle={1\over 2}(\cos^2\varphi \cos 2\omega_+t
+\sin^2\varphi \cos 2\omega_-t + 1).
\end{eqnarray}
Then the Fourier transformation of the time evolution of the mean value $\langle\sigma^x_0(t)\rangle$  has $\delta$-peaks at the frequencies related with the energy levels of $H_T$ and $H$. We obtain
\begin{eqnarray}\nonumber
\sigma^x_0(\omega)
={1\over 2\pi}\int_{-\infty}^{\infty}d t\langle\sigma^x_0(t)\rangle e^{i\omega t}
={1\over 4}\cos^2\varphi (\delta(\omega-2\omega_+)+\delta(\omega+2\omega_+))+\\
+{1\over 4}\sin^2\varphi  (\delta(\omega-2\omega_-)+\delta(\omega+2\omega_-)) +
{1\over 2}\delta(\omega),
\end{eqnarray}
where we use notations $\omega_+=E_+/\hbar$ and $\omega_-=E_-/\hbar$ (for $\hbar=1$ $\omega_{\pm}=E_{\pm}$).

\begin{figure}[!!h]
		\centering
	\includegraphics[scale=0.5]{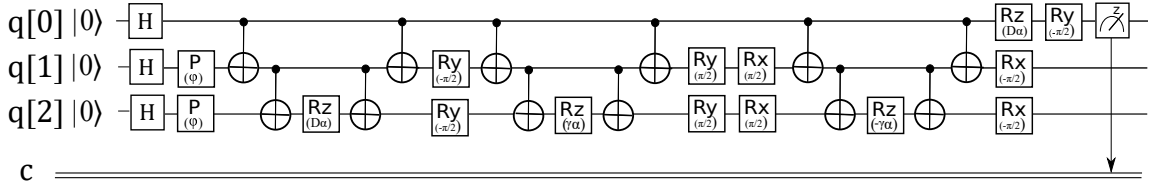}
		\caption{ Quantum protocol for studies of evolution of $\langle\sigma^x_0(t)\rangle$  governed by Hamiltonian (\ref{total}), $\alpha=t$.}
		\label{fig:1}
	\end{figure}

To study $\langle\sigma^x_0(t)\rangle$ on a quantum device we construct quantum protocol presented on Fig. \ref{fig:1}.

In  protocol Fig. \ref{fig:1} the Hadamard gates and the phase shift gates are applied to prepare the initial state (\ref{initial}). We have $\ket{\psi_0}=P_1(\varphi)P_2(\varphi)H_0H_1H_2\ket{000}$.
 To realize the operator of evolution with  Hamiltonian (\ref{total}) we  take into account that with the exactness to the total phase the operator $\exp(-i{D\alpha}\sigma^z_0\sigma^z_1\sigma^z_2/2)$ can be represented as
$CNOT_{01}CNOT_{12}RZ_2(D\alpha)CNOT_{12}CNOT_{01}$. Also,
\begin{eqnarray}
e^{-i{\gamma\alpha\over2}\sigma^z_0\sigma^x_1\sigma^x_2}=e^{-i{\pi\over4}\sigma^y_1}e^{-i{\pi\over4}\sigma^y_2}e^{-i{\gamma\alpha\over2}\sigma^z_0\sigma^z_1\sigma^z_2}e^{i{\pi\over4}\sigma^y_1}e^{i{\pi\over4}\sigma^y_2}.\\
e^{i{\gamma\alpha\over2}\sigma^z_0\sigma^y_1\sigma^y_2}=e^{i{\pi\over4}\sigma^x_1}e^{i{\pi\over4}\sigma^x_2}e^{i{\gamma\alpha\over2}\sigma^z_0\sigma^z_1\sigma^z_2}e^{-i{\pi\over4}\sigma^x_1}e^{-i{\pi\over4}\sigma^x_2}.
\end{eqnarray}
therefore operators $\exp(-i{\gamma\alpha}\sigma^z_0\sigma^x_1\sigma^x_2/2)$, $\exp(i{\gamma\alpha}\sigma^z_0\sigma^y_1\sigma^y_2/2)$  with the exactness to the total phase  can be represented as
\begin{eqnarray} RY_1\left({\pi\over2}\right)RY_2\left({\pi\over2}\right)CNOT_{01}CNOT_{12}RZ_2(\gamma\alpha)CNOT_{12}CNOT_{01}\times\nonumber\\\times RY_1\left(-{\pi\over2}\right)RY_2\left(-{\pi\over2}\right), \\
RX_1\left(-{\pi\over2}\right)RX_2\left(-{\pi\over2}\right)CNOT_{01}CNOT_{12}RZ_2(-\gamma\alpha)CNOT_{12}CNOT_{01}\times\nonumber\\\times RX_1\left({\pi\over2}\right)RX_2\left({\pi\over2}\right),
\end{eqnarray}
respectively. Here $RX_i(\pi/2)$, $RY_i(\pi/2)$  are $X$- and $Z$-rotation gates acting on $q[i]$.
Finally, to find the mean value of $\sigma^x_0$ operator we apply $RY(-\pi/2)$ because operator $\sigma^x_0$ can be represented as $\sigma^x_0=e^{i{\pi\over4}\sigma^y_0}\sigma^z_0e^{-i{\pi\over4}\sigma^y_0}$. So, to calculate the mean value of  $\sigma^x_0$  on the basis of the results of measurement in the standard basis the state of qubit $q[0]$ has to be rotated around the $Y$ axis.

\begin{figure}[h!]
\begin{center}
\subcaptionbox{\label{ff1}}{\includegraphics[scale=0.6, angle=0.0, clip]{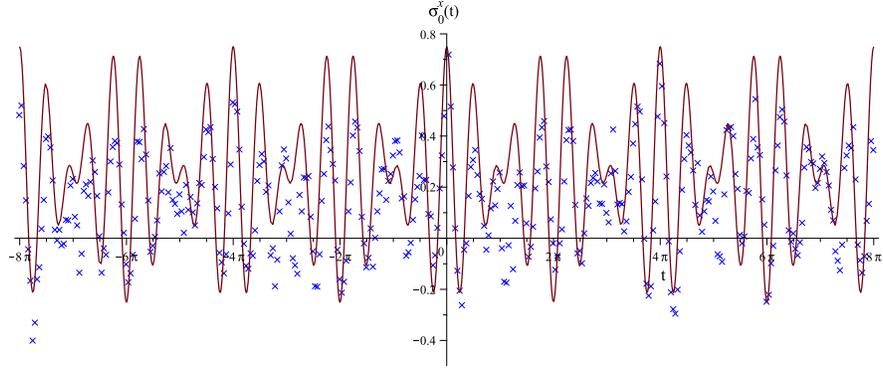}}
\hspace{1cm}
\subcaptionbox{\label{ff1}}{\includegraphics[scale=0.3, angle=0.0, clip]{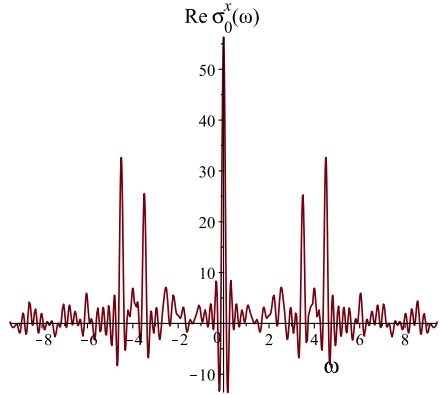}}
\hspace{1cm}
\caption{Evolution of the mean value of $\sigma^x_0$ detected on ibmq-bogota in the case of $D=-2$ and $\gamma=0.25$ (a) and the real part of $\sigma^x_0(\omega)$ found on the basis of results of quantum calculations (b). }
		\label{fig:8}
\end{center}
	\end{figure}

\begin{figure}[h!]
\begin{center}
\subcaptionbox{\label{ff3}}{\includegraphics[scale=0.3, angle=0.0, clip]{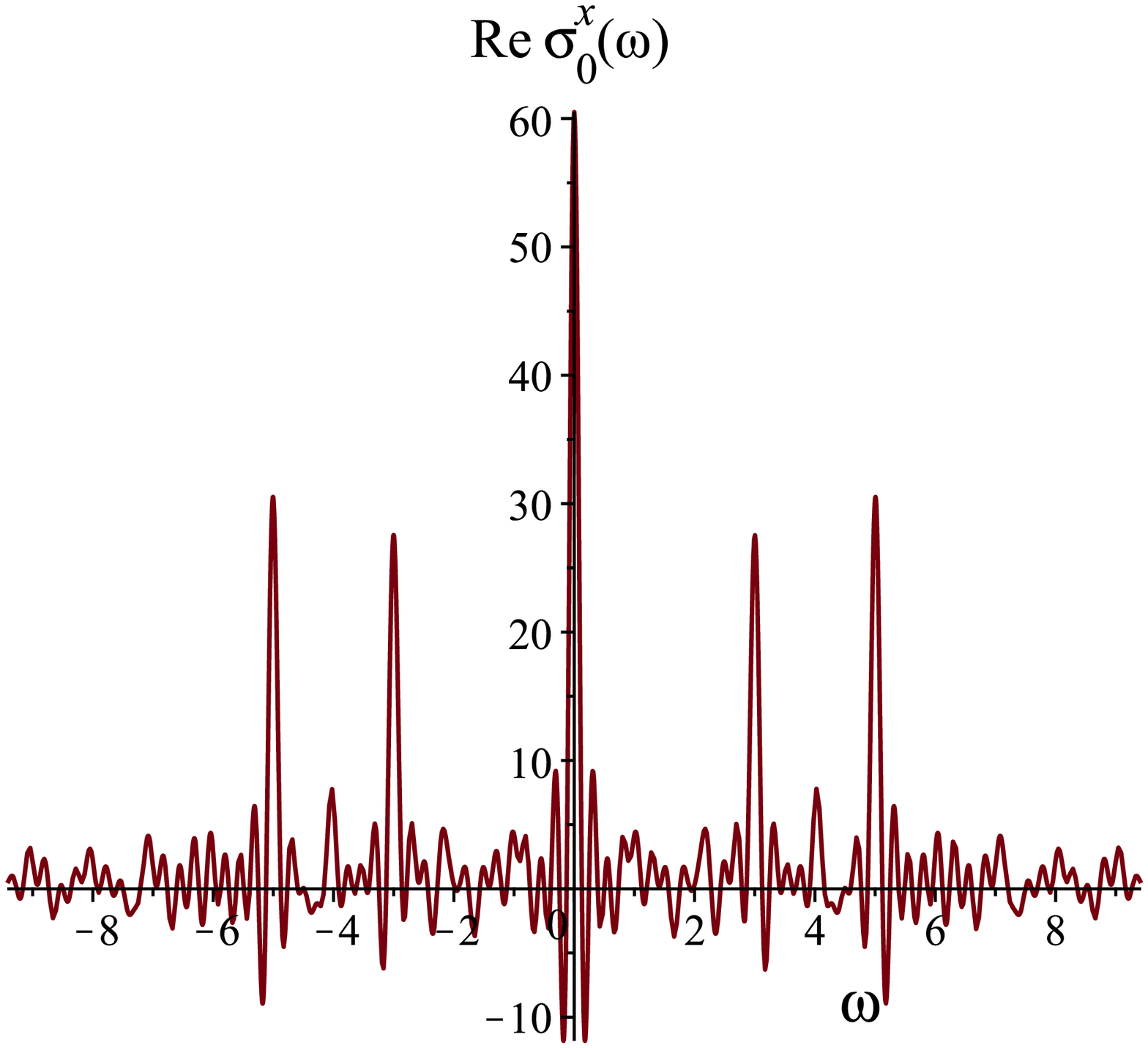}}
\hspace{1cm}
\subcaptionbox{\label{ff2}}{\includegraphics[scale=0.3, angle=0.0, clip]{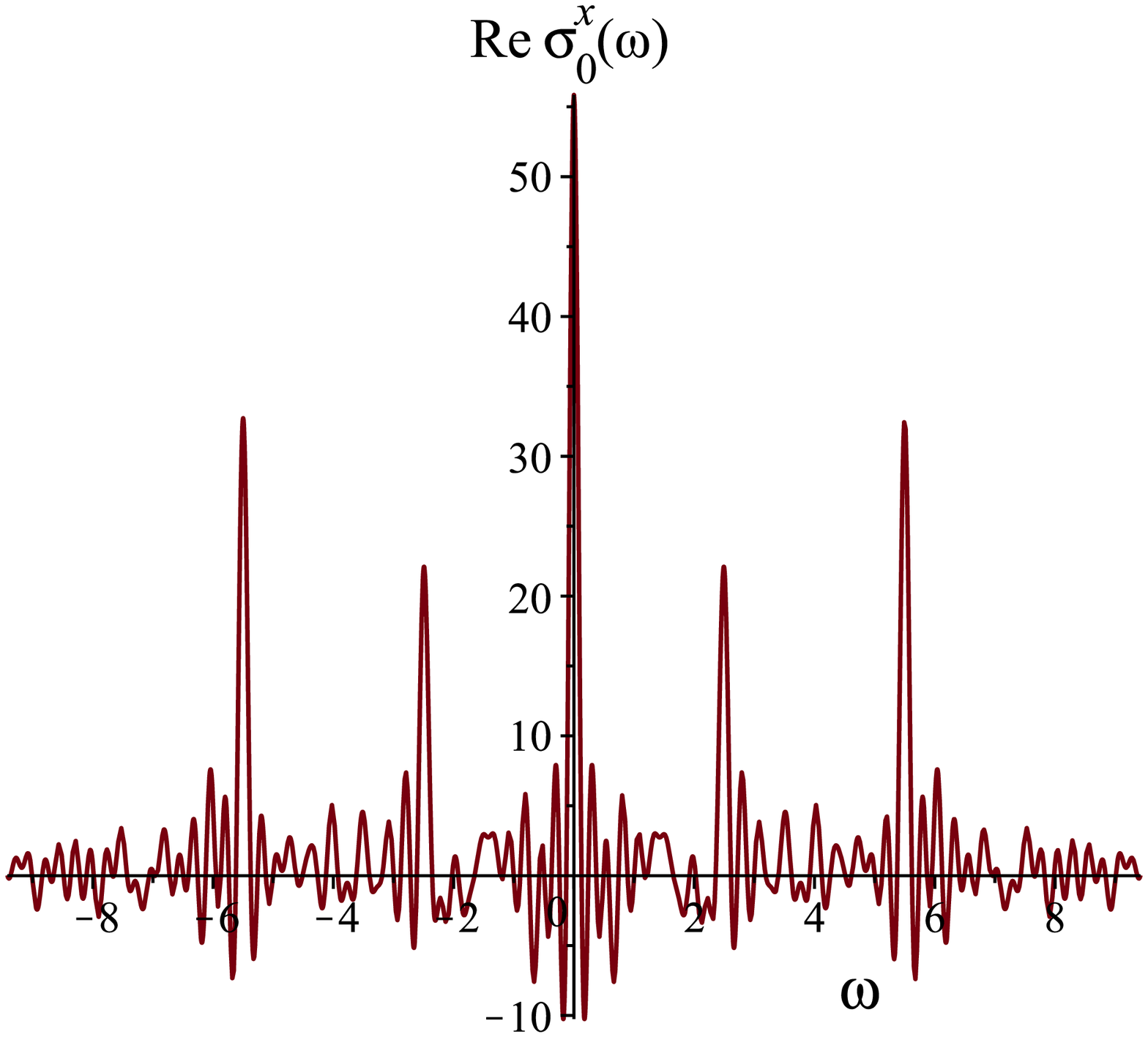}}
\hspace{1cm}
\subcaptionbox{\label{ff3}}{\includegraphics[scale=0.3, angle=0.0, clip]{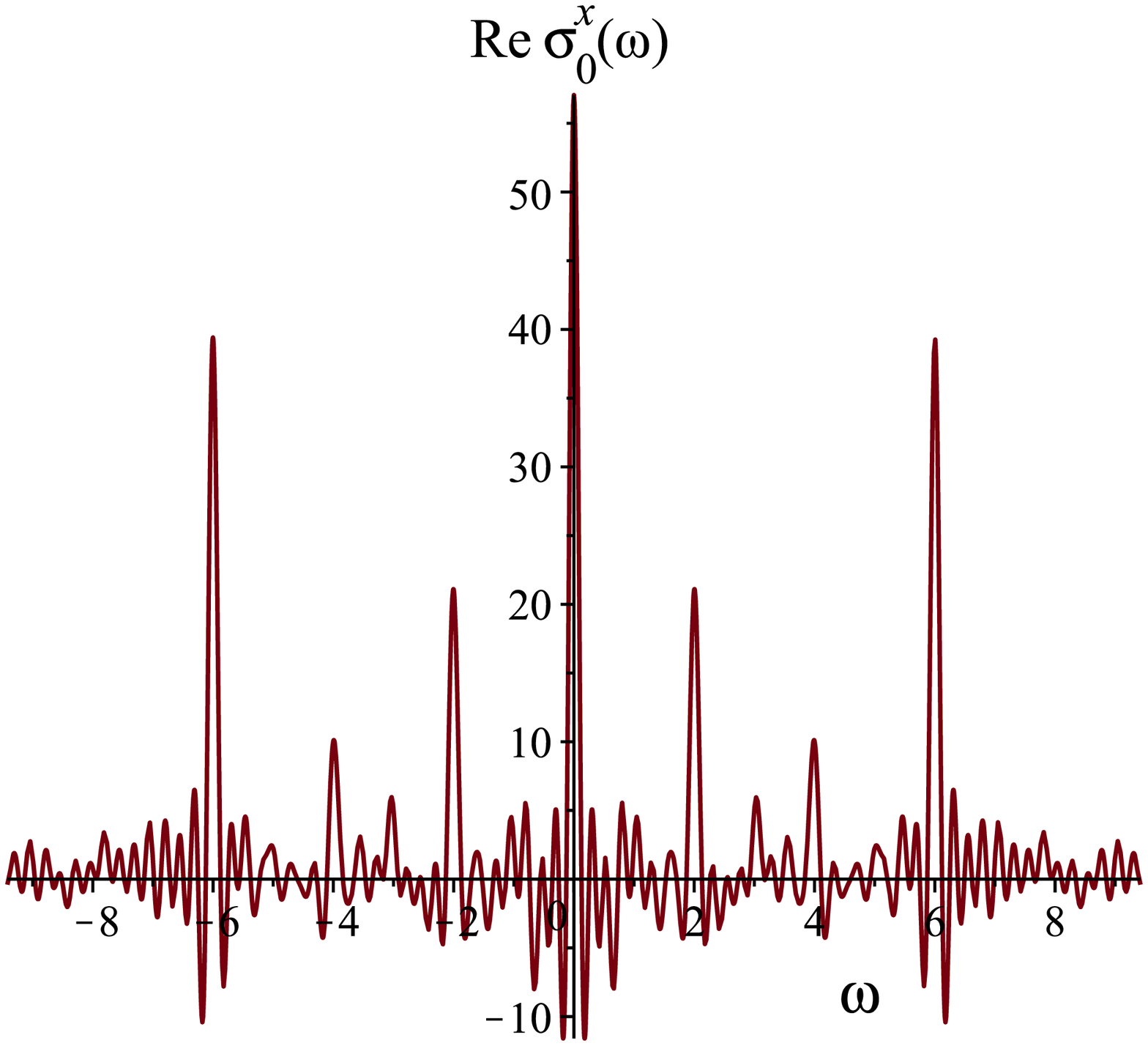}}
\hspace{1cm}
\subcaptionbox{\label{ff1}}{\includegraphics[scale=0.3, angle=0.0, clip]{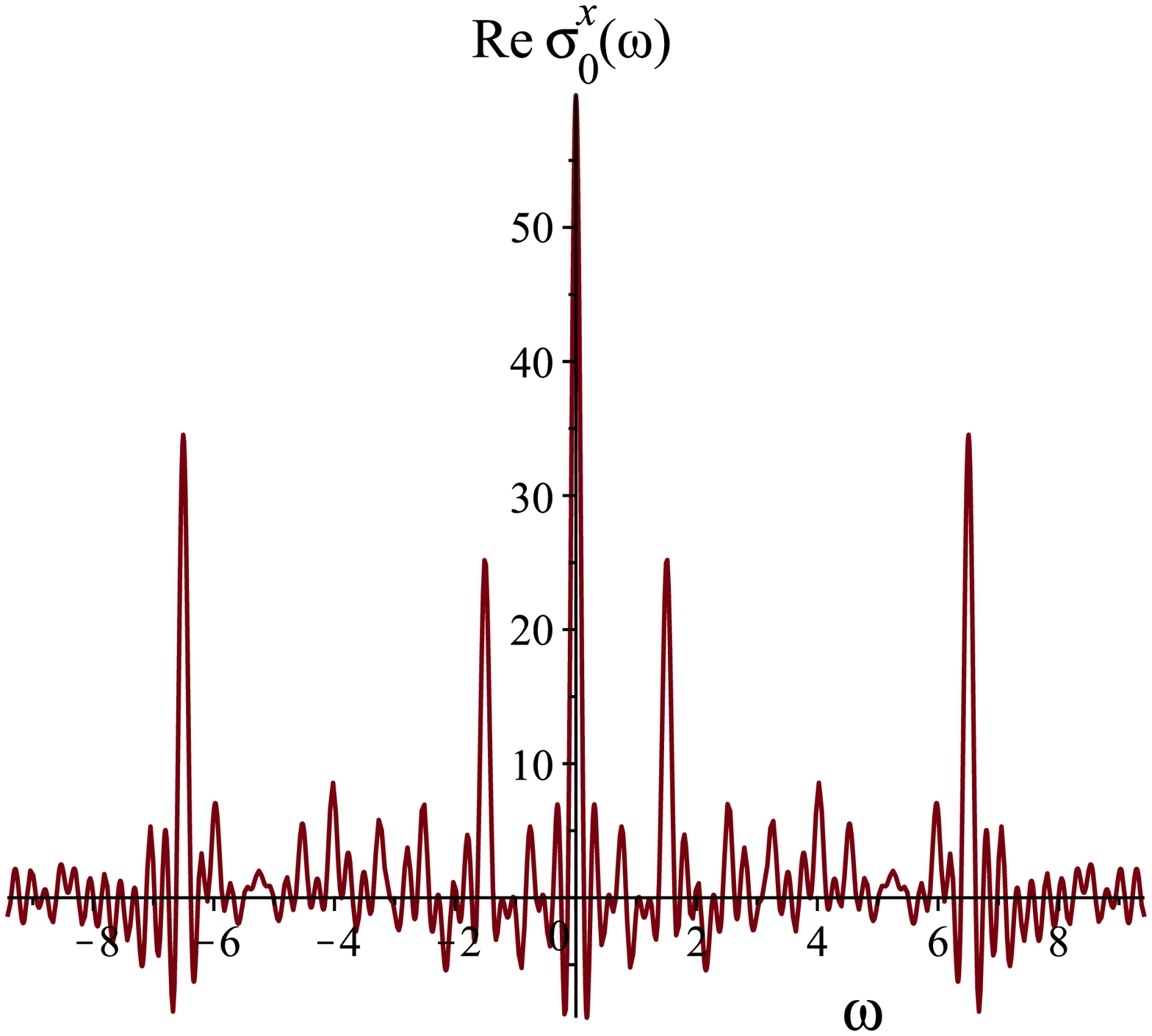}}
\hspace{1cm}
\caption{ The real part of $\sigma^x_0(\omega)$ obtained on the basis of results of quantum calculations for $\braket{\sigma^x_0(t)}$ obtained on ibmq-bogota for $D=-2$ and (a) $\gamma=0.5$, (b) $\gamma=0.75$, (c) $\gamma=1$, (d) $\gamma=1.25$.}
		\label{fig:9}
\end{center}
	\end{figure}

The protocol Fig. \ref{fig:1} was realized on IBM's quantum computer ibmq-bogota for parameter $\alpha$ changing from $-8\pi$ to $8\pi$ with the step $\pi/24$. We choused $\varphi=\pi/4$.  To detect the   energy spectrum splitting we performed quantum calculations for $D=-2$ and different parameters $\gamma$ possessing the following values  $0.25, 0.5, 0.75, 1, 1.25, 1.5, 1.75$.
The results of quantum calculations and the are shown on Figs. \ref{fig:8}, \ref{fig:9}

Note that because of errors of calculations on the quantum device we have also obtained the imaginary part of $\braket{\sigma_0^x (t)}$  that looks as a noise. Therefore,
on Figs. \ref{fig:8}, \ref{fig:9} we plot the real part of  the mean value $\braket{\sigma_0^x (t)}$.

Delta peaks of $\textrm{Re}\,\sigma^x_0(\omega)$ at $\omega_-=-4.5$, $\omega_{+}=-3.5$ correspond to the energies $E_-=-2.25$, $E_+=-1.75$ of (\ref{HS1}) (see Fig. \ref{fig:8} (b)). Similarly
 detecting delta peaks at $\omega_-=-5$, $\omega_{+}=-3$ we obtain  $E_-=-2.5$, $E_+=-1.5$  (see Fig. \ref{fig:9} (a)), delta-peaks at $\omega_-=-5.5$, $\omega_{+}=-2.5$ correspond to $E_-=-2.75$, $E_+=-1.75$ (see Fig. \ref{fig:9} (b)). Obtaining   delta peaks at $\omega_-=-6$, $\omega_{+}=-2$  we quantify  $E_-=-3$, $E_+=-1$ Fig. \ref{fig:9} (c), and delta peaks at $\omega_-=-6.5$, $\omega_{+}=-1.5$  give us energy levels $E_-=-3.25$, $E_+=-0.75$.  So,  we detect  the quantum tunneling splitting energy of the spin $S=1$ on IBM's quantum computer ibmq-bogota. The results of quantum calculations correspond to the analytical ones.

\section{Conclusions}

 Quantum spin-1 tunneling has been observed on IBM's quantum computer, ibmq-bogota. To model spin-1 on the quantum devise we have used its representation by two spins-1/2.

We have proposed to detect spin-1 tunnelling studying time evolution of mean value of z-component of spin-1 operator $\braket{S^z(t)}$. It has been shown that the time dependence of  $\braket{S^z(t)}$ reflects oscillations of spin-1  between two opposite directions  in the result of tunneling.

The  evolution of mean value of z-component of spin-1 operator governed by Hamiltonian describing spin-1 tunnelling has been found (see Fig.  \ref{fig:5}). For this purpose quantum protocol  Fig.  \ref{fig:4} has been realized on the ibmq-bogota. As a  result, the spin-1 tunnelling  between two opposite directions  described by the state vectors $|1\rangle$, $|-1\rangle$  has been observed on a quantum device.

The energy levels of Hamiltonian describing spin-1 tunneling have been quantified on the quantum computer on the basis of studies of a probe spin evolution Figs. \ref{fig:8}, \ref{fig:9}. The splitting of the energy levels of spin-1 has been detected. The results of quantum calculations are in agreement with the theoretical ones.

\section*{Acknowledgments}

This work was supported by Project 2020.02/0196  (No. 0120U104801) from the National Research Foundation of Ukraine.

\end{document}